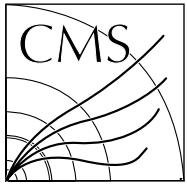

**The Compact Muon Solenoid Experiment**

# CMS Note

Mailing address: CMS CERN, CH-1211 GENEVA 23, Switzerland

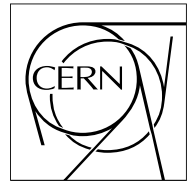

April 12, 2002

# Using Self-Description to Handle Change in Systems

Florida Estrella[1], Zsolt Kovacs[2], Jean-Marie Le Goff[2], Richard McClatchey[1] and Steven Murray[2]

[1]Centre for Complex Cooperative Systems, Univ. West of England, Frenchay, Bristol BS16 1QY UK

[2]CERN, Geneva, Switzerland

*Abstract*

In the web age systems must be flexible, reconfigurable and adaptable in addition to being quick to develop. As a consequence, designing systems to cater for change is becoming not only desirable but required by industry. Allowing systems to be self-describing or description-driven is one way to enable these characteristics. To address the issue of evolvability in designing self-describing systems, this paper proposes a pattern-based, object-oriented, description-driven architecture. The proposed architecture embodies four pillars - first, the adoption of a multi-layered meta-modeling architecture and reflective meta-level architecture, second, the identification of four data modeling relationships that must be made explicit such that they can be examined and modified dynamically, third, the identification of five design patterns which have emerged from practice and have proved essential in providing reusable building blocks for data management, and fourth, the encoding of the structural properties of the five design patterns by means of one pattern, the Graph pattern. In this paper the fundamentals of the description-driven architecture are described – the multi-layered architecture and reflective meta-level architecture, remaining detail can be found in the cited references. A practical example of this architecture is described, demonstrating the use of description-driven data objects in handling system evolution.

Keywords: Database design, systems description, metadata, data modeling

# 1. Introduction

The design and implementation of large-scale, distributed systems has always been a challenging task. Improved software development techniques and design methodologies have somewhat eased this task but have been limited in their scope since they address either specific aspects of implementation strategy or particular views of the design process. With the advent of the Internet and e-commerce, the need for coexistence and interoperation with legacy systems and the demand for reduced 'times-to-market' the need for timely delivery of flexible software has increased. Couple to this the increasing complexity of systems and the requirement for systems to evolve over potentially extended timescales and the importance of clearly defined, extensible models as the basis of rapid systems design becomes a pre-requisite to successful systems implementation.

Many approaches have been proposed to address aspects of design and implementation for modern object-oriented systems. Each has its merits and focusses on concerns such as data modelling, process modelling, state modelling and lifecycle modelling. More or less successful attempts have been made to combine these approaches into modelling languages or methodologies such as OMT [1] and UML [2] but ultimately these lack cohesion since they are often collections of disparate techniques. Recent practitioners' reports on their use have led to proposals for enhancements such as pUML [3] which have recognised and begun to address these failings.

This paper advocates a design and implementation approach which is holistic in nature, viewing the development of modern object-oriented software from a systems standpoint. The design philosophy that is proposed is based on the systematic capture of the description of system elements covering multiple views of the system to be designed (including data, process and time views) using common techniques. The approach advocated here has been termed description-driven systems (otherwise referred to as self-description) and its underlying philosophy is the subject of this paper. Essentially the description-driven approach involves identifying and abstracting the crucial elements (e.g items, processes, lifecycles) of the system under design and creating high-level descriptions of these elements which are stored and managed separately from their instances.

The following section outlines how we have arrived at a description-driven systems philosophy and that is followed by a discussion of the role of reflection in self-description. A later section describes the model which underpins the description-driven systems approach. A practical example of the use of this approach is introduced later in the paper which is brought to a close with conclusions based on the implications of the use of a description-driven systems design approach. This paper advances the work outlined in previous CMS Notes (CMS NOTE 1999/045, CMS NOTE 2001/026 and CMS CR 2001/009).

# 2. Methods, Patterns & Frameworks

Our experience over many years of implementing object-oriented systems has revealed compelling evidence that most effective software reuse comes from the reuse of high level design artifacts. One reason for this is that underlying software technology changes so rapidly, this being especially true in software projects that have long timescales, thereby making code reuse difficult. For example object technology has witnessed, in a short space of time an evolution from languages such as Smalltalk, ADA, C++, Java through middleware such as EJB, COM+, Active X and OMG CORBA and the object community is still in a state of flux.

Where we have experienced more success in reuse of software artifacts is with visual modelling languages such as OMT and UML. The creation and evolution of graphical models using UML has allowed us to specify, visualize, construct and document the artifacts of the software systems we have built, adopting an approach in which we concentrate on the descriptive elements of UML. UML has increasingly become the universal object analysis and design modelling standard and as a consequence of this we have over the years been able to reuse large-grained architectural frameworks and patterns which have been captured in UML.

Design patterns [4] focus on reuse of abstract designs and software architecture, which is usually described using graphical modeling notation. In the object oriented community well known patterns are named, described and cataloged for reuse by the community as a whole. We have not only used many well-known patterns but in the domain of description-driven systems development (see below) we have discovered new patterns. It has enabled us to make use of design patterns that were proven on previous projects and is an example of reuse at the larger grain level. UML diagrams are able to describe pattern structure but provide little support for describing pattern behavior or any notation for the pattern template.

Frameworks are reusable semi-complete applications that can be specialized to produce custom applications [5]. They specify reusable architectures for all or part of a system and may include reusable classes, patterns or templates. Frameworks focus on reuse of concrete design algorithms and implementations in a particular programming language. Frameworks can be viewed as the reification of families of design patterns and are an important step towards the provision of a truly holistic view of systems design

Emerging and future information systems however require a more powerful data model that is sufficiently expressive to capture a broader class of applications. Evidence suggests that the data model must be OO, since that is the model providing most generality [6]. The data model needs to be an open OO model, thereby coping with different domains having different requirements on the data model [7]. Object meta-modeling allows systems to have the ability to model and describe both the static properties of data and their dynamic relationships, and address issues regarding complexity explosion, the need to cope with evolving requirements, and the systematic application of software reuse.

To be able to describe system and data properties, object meta-modeling makes use of meta-data. Meta-data consists of information defining other data. Data descriptions, such as information about how data items are organized and related to one another, and system descriptions, such as information about how components are specified and interrelated, permit run-time access, consequently allowing objects to be composed and managed dynamically. The judicious use of meta-data leads to heterogeneous, extensible and open systems [8][9]. Meta-data make use of a meta-model to describe domains. Meta-modeling [10] creates flexible systems offering the following - reusability, complexity handling, version handling, system evolution and inter-operability. Promotion of reuse, separation of design and implementation and reification are some of the reasons for using meta-models [11]. As such, meta-modeling is a powerful and useful technique in designing domains and developing dynamic systems.

The use of UML, Patterns and Frameworks as design languages and devices has, in our experience, eased difficulties inherent in the timely delivery of large complex object-based systems. However, each approach addresses only part of the overall 'design space' and fails to enable a holistic view on the design process. In particular they do not easily model the description aspects or meta information emerging from systems design. In other words, these approaches can locate individual pieces in the overall design puzzle but do not enable the overall puzzle to be viewed. What is required is a philosophy which opens up the structure of the puzzle enabling us to answer questions such as 'what is the most appropriate strategy for piecing together the overall puzzle' ? In the next section we look at reflection as the mechanism to open up the design puzzle.

## 3. Reflection in Systems Design

A crucial factor in the creation of flexible information systems dealing with changing requirements is the suitability of the underlying technology in allowing the evolution of the system. Exposing the internal system architecture opens up the architecture, consequently allowing application programs to inspect and alter implicit system aspects. These implicit system elements can serve as the basis for changes and extensions to the system. Making these internal structures explicit allows them to be subject to scrutiny and interrogation.

A reflective system utilizes an open architecture where implicit system aspects are promoted to become explicit first-class *meta-objects* [12] i.e. are *reified*. The advantage of reifying system descriptions as objects is that operations can be carried out on them, like composing and editing, storing and retrieving, organizing and reading. Since these meta-objects can represent system descriptions, their manipulation can result in change in the overall system behaviour. As such, reified system descriptions are mechanisms which can lead to dynamically modifiable systems. Meta-objects, as used here, are the self-representations of systems describing how the system's internal elements can be accessed and manipulated. These self-representations are causally connected to the internal structures they represent i.e. changes to these self-representations immediately affect the underlying system. The ability to dynamically augment, extend and redefine system specifications can result in a considerable improvement in flexibility. This leads to dynamically modifiable systems which can adapt and cope with evolving requirements.

There are a number of OO design techniques which encourage the design and development of reusable objects. In particular design patterns are useful for creating reusable OO designs [13]. Design patterns for structural, behavioral and architectural modeling have been documented and have provided software engineers rules and guidelines which they can immediately (re-)use in software development. Reflective architectures that can dynamically adapt to new user requirements by storing descriptive information which can be interpreted at runtime have lead to so-called Adaptive Object Models [14], [15]. These are models which provide meta-information about domains that can be changed on-the-fly. Such an approach, proposed by Yoder, is very similar to the approach adopted in this paper.

In OO programming, the class of a class object is a meta-class. Meta-objects, therefore, are implemented as meta-classes. Object models used in most class-based programming language are fixed and closed. These object models do not allow the introduction and extension of modelling primitives to cater for specific application needs. The concept of meta-classes is a key design technique in improving the reusability and extensibility of these languages. A Description-Driven System (DDS) [10] architecture, as advocated in this paper, is an example of a reflective meta-layer (i.e. meta-level and multi-layered) architecture. It makes use of meta-objects to store domain-specific system descriptions, which control and manage the life-cycles of meta-object instances i.e. domain objects. The separation of descriptions from their instances allows them to be specified and managed and to evolve independently and asynchronously. This separation is essential in handling the complexity issues facing many CMS computing applications and allows the realization of inter-operability, reusability and system evolution as it gives a clear boundary between the application's basic functionalities from its representations and controls. As objects, reified system descriptions of DDSs can be organized into libraries or frameworks dedicated to the modelling of languages in general, and to customizing its use for specific domains in particular.

## 4. Description-Driven Systems

In modeling complex information systems, it has been shown that at least four modeling layers are required [16]. Each layer provides a service to the layer above it and serves as a client to the layer below it. The meta-meta-model layer defines the language for specifying meta-models. Typically more compact than the meta-model it describes, a meta-meta-model defines a model at a higher level of abstraction than a meta-model. Elements of the meta-meta-model layer are called meta-meta-objects. Examples of meta-meta-objects include MetaClass, MetaAttribute and MetaAssociation. These meta-meta-objects are also meta-classes whose instances are constructs corresponding to meta-model constructs.

The meta-model layer defines the language for specifying models. A meta-model is an instance of a meta-meta-model. It is also more elaborate than the meta-meta-model that describes it. Elements of the meta-model layer are called meta-objects, examples of which include Class, Attribute and Association. The model layer defines the language for specifying information domains. In this case, a model is an instance of a meta-model. Elements like Student, Teacher and Course classes are domain-specific examples of elements of the model layer. The bottom layer contains user objects and user data. The instance layer describes a specific information domain. Domain examples of user objects include the instances of Student, Teacher and Course classes. The Object Management Group (OMG) [17] standards group uses the same architecture based on model abstraction, with the Meta-Object Facility (MOF) model and the Unified Modeling Language (UML) [2] model defining the language, for the meta-meta-model and meta-model layers, respectively.

Orthogonal to the model abstraction inherent in multi-layered meta-modeling approach is the information abstraction which separates *descriptive* information from the data they are describing. These system descriptions are called *meta-data*, as they are information defining other data. A reflective open architecture typifies this abstraction. A reflective open architecture is divided into two levels - the meta-level where the descriptive information reside and the base-level which stores the application data described by the meta-level elements. The meta-level contains the meta-data objects (also referred to as meta-objects in this paper) which hold the meta-data. These meta-objects manage the base-level objects.

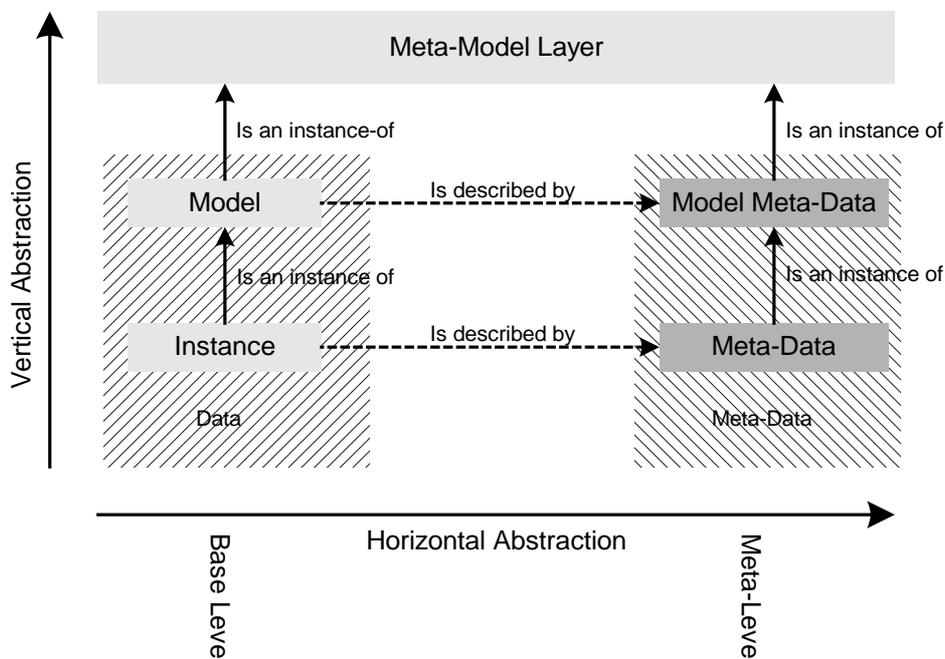

**Figure 1 : A description-driven architecture**

In a description-driven system as we define it, descriptions are separated from their instances and managed independently to allow the descriptions to be specified and to evolve asynchronously from particular instantiations of those descriptions. Separating descriptions from their instantiations allows new versions of elements (or element descriptions) to coexist with older versions. This paper proposes an architecture which combines a multi-layered meta-modeling approach with a meta-level architecture [18]. The description-driven architecture is illustrated in Figure 1. It is the combination of a multi-layered architecture (based on an *Instance of* relationship) and of a meta-level architecture based on the *Is described by* relationship) that constitutes this description-driven systems architecture.

The layered architecture on the left hand side of figure 1 is typical of layered systems and the multi-layered architecture specification of the OMG. The relationship between the layers is *Is an instance of*. The instance layer contains data which are instances of the domain model in the model layer. Similarly, the model layer is an instance of the meta-model layer. On the right hand side of the diagram is another form of model abstraction. It shows the increasing abstraction of information from meta-data to model meta-data, where the relationship between the two is Is an instance of as well. These two architectures provide layering and hierarchy based on abstraction of data and information models.

The horizontal view in figure 1 provides an alternative abstraction where the relationship of meta-data and the data they describe are made explicit. This view is representative of the information abstraction and the meta-level architecture discussed earlier. The meta-level architecture is a mechanism for relating data to information describing data, where the link between the two is *Is described by*. As a consequence, the dynamic creation and specification of object types is promoted.

The separation of system type descriptions from their instantiations allows the asynchronous specification and evolution of system objects from system types, consequently, descriptions and their instances are managed independently and explicitly. The dynamic configuration (and re-configuration) of data and meta-data is useful for systems whose data requirements are unknown at development time. It is the combination of the *instance of* and *is described by* relationships that provides the holistic approach inherent to description-driven systems.

## 5. A Relational Description-Driven System

Paper [19] discusses an analogous modeling architecture which is based on relational models. The horizontal abstraction of the architecture is also based on Instance-of relationship and a meta-modeling approach. Figure 2, taken from [19], illustrates a relational equivalent of the DDS architecture. Meta-data are generated and extracted from data elements' structures. The data layer corresponds to the instances that are manipulated in the system. The type of meta-data that can be extracted in the data layer concerns the physical aspects of the data such as the volume or localization. The model layer corresponds to concepts for model description describing data and meta-data in the data layer. In relational models, the model layer is the application model. The meta-data that can be extracted from the model layer are descriptive information concerning the physical structure, e.g. data dictionary.

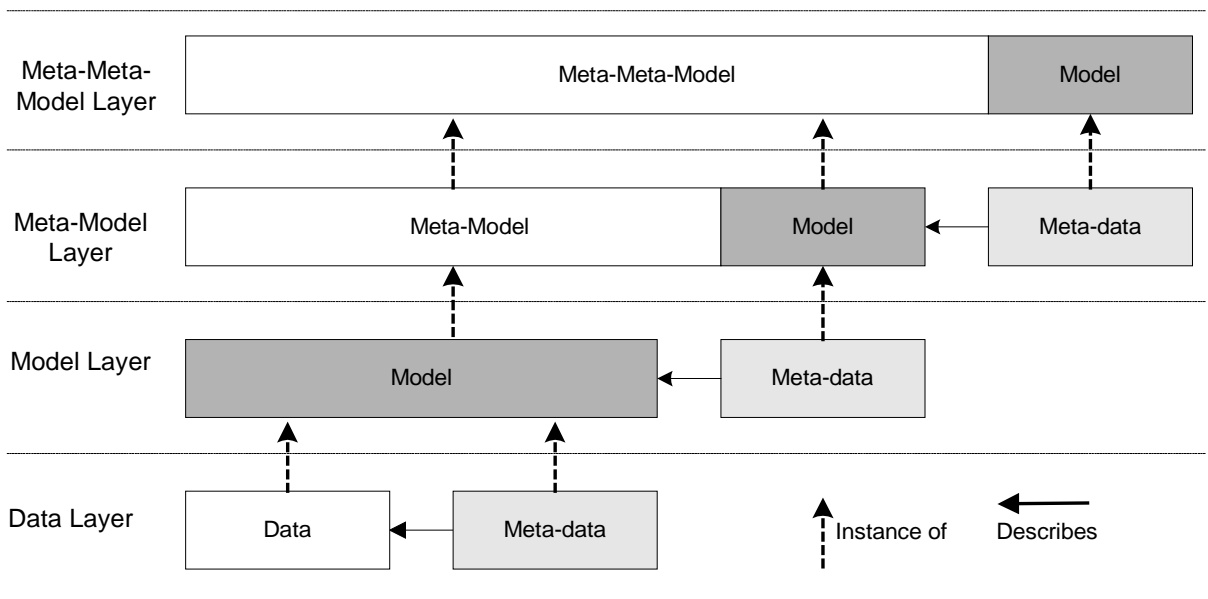

**Figure 2 : A relational equivalent of the Description-Driven system architecture.**

The meta-model layer defines the model formalism that is used in the system. As the architecture is intended for relational systems, the meta-model layer describes the concepts of the relational model, e.g. Table, Relation and Key and the model relating these primitives together. At this level, the type of meta-data that can be extracted concerns tools and methods for inter-operating and relating the different relational models. The meta-meta-model layer is the root modeling level which supports inter-operability and extensibility. The meta-meta-model layer uses conceptual graphs [20] to allow homogeneous representation and manipulation of the lower levels. Conceptual graphs are a formalism where the universe of discourse can be modeled by concepts and conceptual

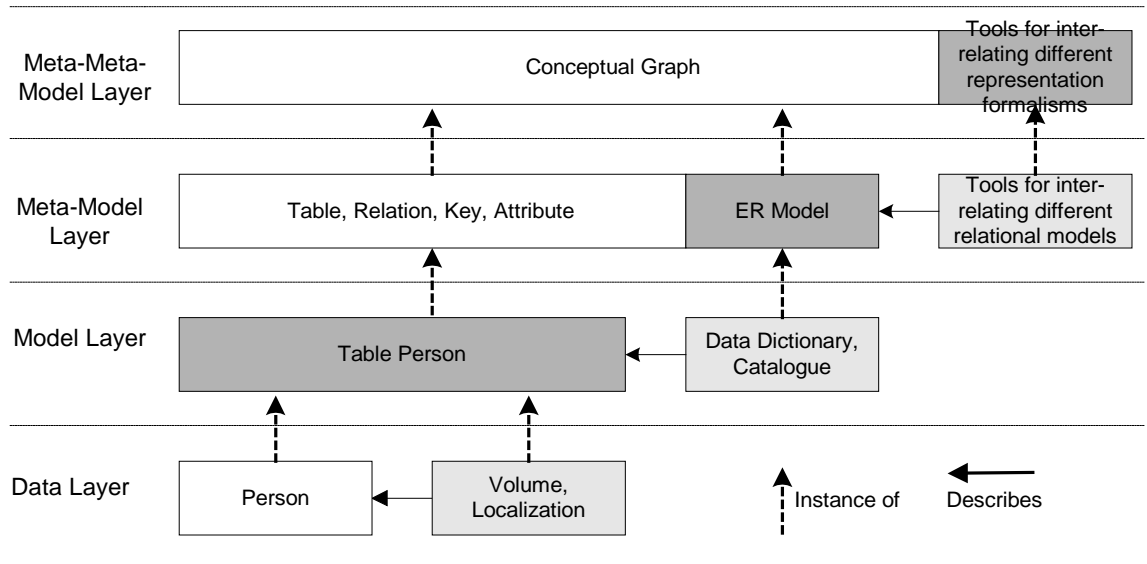

relations. A relational architecture example is shown in Figure 3.

**Figure 3 : A relational architecture example.**

Our proposed DDS architecture is similar to the architecture in [19]. Both horizontal layerings are based on model abstraction, and meta-data are used as descriptive information to describe the concepts for managing data at each layer. The two architectures are orthogonal in the modeling paradigm used - relational model for [19] and object model for DDS. Thus, the relational architecture uses conceptual graphs and the DDS architecture uses the MOF as its representation formalism in the meta-meta-model layer. In the meta-model layer, the relational architecture uses a relational model and the DDS architecture uses the UML.

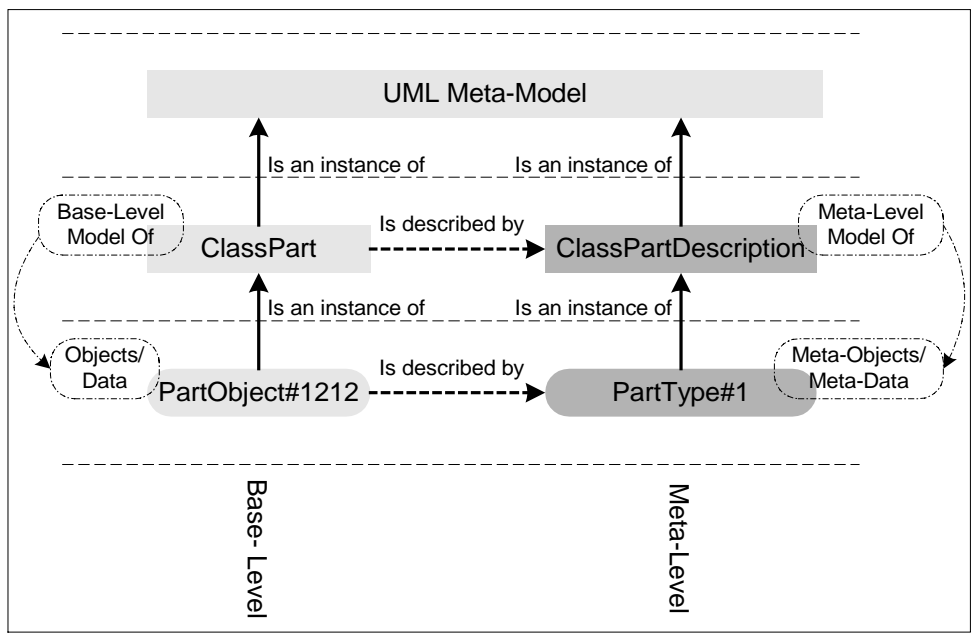

**Figure 4 : The Description-Driven architecture of CRISTAL**

# 6. CRISTAL as a Description-Driven System

The design of the CRISTAL prototype was dictated by the CMS requirements for adaptability over extended timescales, for system evolution, for interoperability, for complexity handling and for reusability. In adopting a description-driven design approach to address these requirements, the separation of object instances from object descriptions instances was needed. This abstraction resulted in the delivery of a three layer description-driven architecture. The model abstraction (of instance layer, model layer, meta-model layer) has been adopted from the OMG MOF specification [20], and the need to provide descriptive information, i.e. meta-data, has been identified to address the issues of adaptability, complexity handling and evolvability.

Figure 4 illustrates the CRISTAL architecture. The CRISTAL model layer is comprised of class specifications for CRISTAL type descriptions (e.g. PartDescription) and class specifications for CRISTAL classes (e.g. Part). The instance layer is comprised of object instances of these classes (e.g. PartType#1 for PartDescription and Part#1212 for Part). The model and instance layer abstraction is based on model abstraction and Is an instance of relationships. The abstraction based on meta-data abstraction and Is described by relationship leads to two levels - the meta-level and the base-level. The meta-level is comprised of meta-objects and the meta-level model which defines them (e.g. PartDescription is the meta-level model of PartType#1 meta-object). The base-level is comprised of base objects and the base-level model which defines them (e.g. Part is the base-level model of the Part#1212 object).

In the CMS experiment, production models change over time. Detector parts of different model versions must be handled over time and coexist with other parts of different model versions. Separating details of model types from the details of single parts allows the model type versions to be specified and managed independently, asynchronously and explicity from single parts. Moreover, in capturing descriptions separate from their instantiations, system evolution can be catered for while production is underway and therefore provide continuity in the production process and for design changes to be reflected quickly into production. As the CMS construction is one-of-a-kind, the evolution of descriptions must be catered for. The approach of reifying a set of simple design patterns as the basis of the description-driven architecture for CRISTAL has provided the capability of catering for the evolution of a rapidly changing research data model. In the two years of operation of CRISTAL it has gathered over 20 Gbytes of data and been able to cope with 25 evolutions of the underlying data schema without code or schema recompilations.

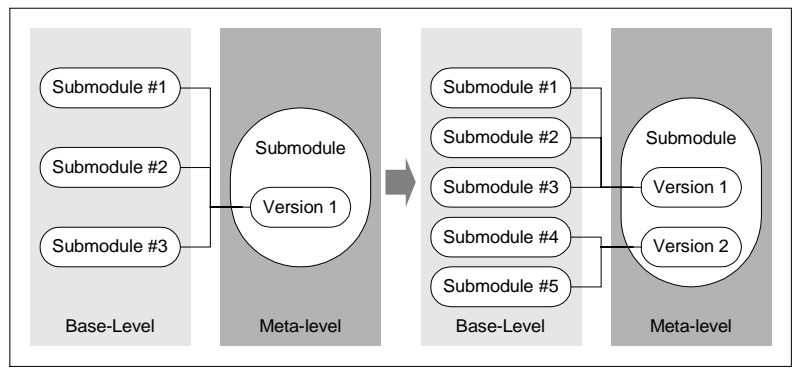

**Figure 5: Evolving CMS Descriptions**

The evolving CMS descriptions are illustrated in Figure 5. Submodule Version 1 is a type object describing physical detector parts Submodule#1, Submodule#2 and Sub-module#3. Submodule Version 2 (another type object) is a new version of the same type specification, and coexists with Submodule Version 1 and its instances. In this example, Submodule Version 1 and Submodule Version 2 are instances of ClassPart-Description (in the meta-level). As new instances can be dynamically added into the system, consequently new versions and new type objects are handled transparently and automatically. In the base-level, Submodule#1 is an instance of ClassPart. The Is described by relationship between the meta-level and the base-level elements allows for instantiations of physical parts to be described by versions of type objects in the meta-level. Hence, the separation of type descriptions in the meta-level from the data objects they describe caters for evolving system specifications.

# 7. Conclusions

Reflection provides a sound foundation for the specification of meta-level architectures. Reflection gives the capability of customizing system behavior through explicit descriptive mechanisms of implicit system aspects. The transformation of implicit system aspects to explicit description meta-objects is termed reification. These description meta-objects comprise the meta-layer which manages and controls the life cycles of the base-layer objects it describes. By reifying system descriptions as meta-objects, they can be treated, accessed and altered as objects. As a consequence, this work has demonstrated that through reification, system behavior can be accessed and manipulated at run-time thereby creating a dynamically modifiable system.

The combination of a multi-layered meta-modeling architecture and a reflective meta-level architecture resulted in what has been referred to in this paper as a description-driven systems (DDS) architecture. A DDS architecture, is an example of a reflective meta-layer architecture. The CRISTAL DDS architecture was shown to have two abstractions. The vertical abstraction is based on the OMG meta-modeling standard, and has three layers - instance layer, model layer and meta-model layer. This paper has proposed an orthogonal horizontal abstraction mechanism which complements the OMG approach. The horizontal abstraction is based on the meta-level architecture approach, and has two layers - meta-level and base-level. The relationship between the vertical layers is Instance-of and the relationship between the horizontal layers is Describes.

The approach taken in this paper is consistent with the OMG's goal of providing reusable, easily integrated, easy to use, scalable and extensible components [21][22]. The Common Warehouse Metamodel (CWM) specification [23] has been recently adopted by OMG. The CWM enables companies better to manage their enterprise data, and makes use of UML, XML and MOF. The specification provides a common meta-model for warehousing and acts as a standard translation for structured and unstructured data in enterprise repositories, irrespective of proprietary database platforms.

Likewise, and as noted earlier, the contributions of this work complement the ongoing research on Active Object Model (AOM) espoused in [24][25]. A system with an AOM (also called Dynamic Object Model) has an explicit object model which is stored in the database, and interpreted at run-time. Objects are generated dynamically from the schema meta-data which represent data descriptions. The AOM approach uses reflection in reifying implicit data aspects (e.g. database schema, data structures, maps of layouts of data objects, references to methods or code).

The ubiquity of change in current information systems have contributed to the renewed interest in improving underlying system design and architecture. Reflection, meta-architectures and layered systems are the main concepts this paper has explored in providing a description-driven architecture which can cope with the growing needs of many computing environments. The description-driven architecture has two orthogonal abstractions combining multi-layered meta-modeling with open architectural approach allowing for the separation of description meta-data from the system aspects they represent.

The description-driven philosophy facilitated the design and implementation of the CRISTAL project which required mechanisms for handling and managing evolving system requirements. In conclusion, it is interesting to note that the OMG has recently announced the so-called Model Driven Architecture as the basis of future systems integration [29]. Such a philosophy is directly equivalent to that expounded in this and earlier papers on the CRISTAL description-driven architecture.

## Acknowledgments

The authors take this opportunity to acknowledge the support of their home insti-tutes. A. Bazan, N. Baker, A.Branson, P. Brooks, G. Chevenier, T. Le Flour, C. Koch, S. Lieunard and N. Toth are thanked for their assistance in developing the CRISTAL software.

## References

1. J.Rumbaugh et al., "Object-Oriented Modeling & Design"  Prentice Hall (1991)
2. "The Unified Modeling Language (UML) Specification", URL http://www.omg.org/technology/uml/
3. "pUML Initial Submission to OMG's RFP for UML 2.0 Infrastructure". URL http://www.cs.york.ac.uk/puml/
4. F. Buschmann et al., "Pattern-Oriented Software Architecture: A System of Patterns". Wiley, N.Y. 1996.
5. M. Fayad et al. "Building Application Framework". Wiley, N.Y., 1999
6. T. Ozsu and P. Valduriez, "Principles of Distributed Database Systems", Second Edition, Prentice Hall International, 1999.
7. W. Klas and M. Schrefl, "Metaclasses and their Application. Data Model Tailoring and Database Integration", Lecture Notes in Computer Science 943. Springer. 1995.
8. S. Crawley, et. al., "Meta Information Management", Proceedings of the Second IFIP International Conference on Formal Methods for Open Object-based Distributed Systems, Canterbury, United Kingdom, July 1997.
9. S. Crawley, et. al., "Meta-meta is Better-better!", Proceedings of the IFIP WG 6.1 International Conference on Distributed Applications and Interoperable Systems, 1997.
10. Z. Kovacs, "The Integration of Product Data with Workflow Management Systems", PhD Thesis, University of West of England, Bristol, England, April 1999.
11. M. Blaha, W. Premerlani, "Object-Oriented Modeling and Design for Database Applications", Prentice Hall, 1998.
12. G. Kiczales, "Metaobject Protocols: Why We Want Them and What Else Can They Do?", Chapter in Object-Oriented Programming: The CLOS Perspective, pp 101-118, MIT Press, 1993.

13. E. Gamma, R. Helm, R. Johnson and J. Vlissides, "Design Patterns: Elements of Reusable Object-Oriented Software", Addison-Wesley, 1995.
14. B. Foote and J. Yoder., "Meta-data and Active Object-Models". Proc. of the Int. Conference on Pattern Languages Of Programs, Monticello, Illinois, USA, August 1998.
15. J. Yoder, F. Balaguer & R. Johnson., "Architecture and Design of Adaptive Object-Models". Proc of OOPSLA 2001, Intriguing Technology Talk, Tampa, Florida. October 2001.
16. M. Staudt, A. Vaduva and T. Vetterli, "Metadata Management and Data Warehousing", Technical Report 21, Swiss Life, Information Systems Research, July 1999.
17. The Object Management Group (OMG), URL http://www.omg.org.
18. F. Estrella, "Objects, Patterns and Descriptions in Data Management", PhD Thesis, Uni-versity of the West of England, Bristol, England, December 2000.
19. B. Kerherve and O. Gerbe, "Models for Metadata or Metamodels for Data", Proceedings of the Second IEEE Metadata Conference, Maryland, USA, September, 1997.
20. "The Meta- Object Facility (MOF) Specification", URL http://www.dstc.edu.au/Products/CORBA/MOF/.
21. "E-Business: Success on a Sound Software Architecture", OMG White Paper, Software Magazine, June 2000.
22. J. Waters, "OMG Produces Spec for Warehouse Meta Data", ITworld.com, July 2000
23. L. Morgan, "OMG Adopts CWM Specification", Software Development (SD) Times, July 2000.
24. J. Yoder and B. Foote, "Evolution, Architecture and Metamorphosis", in Pattern Languages of Program Design 2, Addison-Wesley, 1996. Originally presented at the Second Conference on Patterns Languages of Programs (PLoP), 1995.
25. OMG Publications., "Model Driven Architectures - The Architecture of Choice for a Changing World". See http://www.omg.org/mda/index.htm